\documentstyle[sprocl,epsf,epsfig,wrapfig]{article}

\def\Journal#1#2#3#4{{#1} {\bf #2}, #3 (#4)}

\def\NCA{\em Nuovo Cimento}

\def\NPB{{\em Nucl. Phys.} B}
\def\PLB{{\em Phys. Lett.}  B}
\def\PRL{\em Phys. Rev. Lett.}
\def\PRD{{\em Phys. Rev.} D}
\def\ZPC{{\em Z. Phys.} C}
\def\ZPA{{\em Z. Phys.} A}

\def\be{\begin{equation}}
\def\ee{\end{equation}}
\def\bea{\begin{eqnarray}}
\def\eea{\end{eqnarray}}

\begin{document}

\title{ SPIN-DEPENDENT FRAGMENTATION FUNCTIONS FOR BARYONS
IN A DIQUARK MODEL}                 

\author{ ANATOLY D. ADAMOV, GARY R. GOLDSTEIN }

\address{Department of Physics, Tufts University, Medford,\\ MA 02155, USA}

\maketitle\abstracts{
The perturbative QCD calculation of heavy quark fragmentation into a heavy 
meson is extended to predict fragmentation into heavy flavor baryons. This 
is accomplished by implementing the quark-diquark model of the baryons. 
Several diquark form factors are used to enable the integration over the
virtual heavy quark momentum.
The resulting spin independent functions for charmed quarks to fragment
into charmed baryons with 
spin 1/2 and 3/2 are compared with recent data. 
Predictions are made for the spin dependent fragmentation functions as 
well, particularly for the functions $\hat{g}_1$ in the 
case of spin 1/2 baryons.}

\section{Introduction}

Fragmentation functions have received considerable attention in recent
years. While experimental information beyond the pion distribution
(presumably from light quarks) has been slow in accumulating, theoretical
interest has been growing. The particular functional form for heavy
flavored quarks to fragment into heavy mesons has been studied using
Operator Product Expansion techniques, light cone quantization, QCD
perturbation theory, Heavy Quark Effective Theory, and other methods.
Some of these methods yield general properties that reflect the overall
structure of QCD, as it is currently understood. Other approaches take
particular models of the low energy behavior expected from QCD, but in
regions that are not perturbatively calculable. The particulars of the
various approaches will be subject to some experimental scrutiny in the
future, but are not yet put to the test. One general feature is known - 
the peak of the hadron distribution moves toward higher momenta as the 
quark mass increases. This feature is a result of the kinematics implicit 
in most models of the non-perturbative process, and is incorporated in the 
phenomenological Peterson function\cite{peterson} that is used to fit 
the sparse data on heavy quark fragmentation\cite{CLEO}.

Even more difficult to test experimentally are the spin dependences of the
fragmentation processes. Yet these dependences are important to know. They
reflect the details of the primarily non-perturbative mechanism by which parton
polarization is passed on to the hadrons. In this sense, the
spin-dependent fragmentation involves the reverse of the process by which
the nucleon spin is shared by its partons (the``spin crisis''), and may
reveal a similarly mysterious decoupling of valence quark spin and hadron
spin for some regions of kinematics. 

Over the last several years a number of theorists have noticed that for
fragmentation of a heavy flavor quark into ``doubly heavy'' mesons, like
the
c-quark into the $J/\Psi$ or the b-quark into the $B_c$, perturbative QCD
may be applicable\cite{chang,braaten}. If this is the case, the 
fragmentation functions are
calculable, at an appropriate scale, and QCD radiative corrections can be
obtained from the Renormalization Group or the Altarelli-Parisi equations.
Such calculations have been performed and scrutinized. It has been shown
that in the heavy quark limit (i.e. the mass goes to infinity) the
functions have the form expected from more general 
considerations\cite{randall}. This
corresponds to the heavy meson taking all of the heavy quark's momentum;
the distribution becomes a delta function at $z=1$. The $1/m_Q$ 
corrections are calculated also\footnote{It is not clear to the present authors 
that all sources for 
such terms have been considered, particularly corrections coming from the 
treatment of the Bethe-Salpeter bound state wavefunction for the meson}. In
any case, this approach can predict the spin-dependent fragmentation
functions along with their momentum and mass dependences. In the heavy 
mass limit, of course, the spin of the heavy quark is conserved, so the 
spin dependence is simple. What is of phenomenological interest is the 
next order correction, at least, since that has non-trivial spin 
dependence.

In some circumstances the spin dependence of fragmentation is most readily 
studied experimentally by observing baryons rather than mesons. This is 
true for the 
production of hyperons or heavy hyperons ($\Lambda_c$, $\Lambda_b$, etc.), 
wherein the weak, parity violating decays provide polarization 
analyses\cite{chen}. 
To consider fragmentation into baryons in this perturbative scheme, the 
three quark system has to be confronted. A simple alternative is to 
consider the baryons as quark-diquark bound states, and to use the same 
perturbative method as for the mesons. In order for the perturbative 
calculation to be useful the creation of a heavy pair of quarks or 
diquarks must be an intermediate step. Hence, doubly heavy baryon 
fragmentation is an appropriate testing ground for these ideas. It is not 
expected that sufficient data to study this process will be available in 
the near future, however. 

To begin to see the structure it will be 
worthwhile to stretch the region of applicability to the ``singly'' heavy 
baryons. We have been carrying out this program to see the expected spin 
and kinematic dependences, with the hope of providing an experimentally 
testable model. Of immediate interest is the question of whether the 
baryon fragmentation functions have the same kinematic dependence as the 
meson case. In general the answer is no in this model, but the details
have to be studied. Furthermore, the spin dependent fragmentation is 
interestingly distinct from the naive heavy quark limit. The perturbative 
calculation and its results will be presented below, along with a 
comparison with some recent data.

\section{Perturbative calculation}

The first calculations of the fragmentation functions in the perturbative 
scheme were applied to some of the inclusive heavy flavor meson decays of 
the $Z^0$, as produced 
at LEP. The partial width for this inclusive process can be written
in general for a hadron H as
\begin{equation}
d\Gamma(Z^0 \rightarrow H(E) + X) = \sum_i\!\int_0^1\!dz\,d\hat{\Gamma}\!
(Z^0 \rightarrow i(E/z) + X,\mu)\, D_{i\rightarrow H}\!(z,\mu),
\label{eq:general}
\end{equation}
where H is the hadron of energy E and longitudinal momentum fraction z 
relative to the parton {\it i}, while $\mu$ is the arbitrary scale whose 
value will be chosen to avoid large logarithms.
The fragmentation function enters here in a factorized form (that can be 
maintained through the evolution equations).

Now, consider the final state with one heavy flavor meson, say the $B_c$ 
for definiteness.
To leading order the $B_c$ meson arises from the production of a pair of 
b-quarks, in which one of the quarks fragments into the meson. As Fig.1 
illustrates, with $Q=b, Q'=c, \bar{Q}'=\bar{c}$, the perturbative
contribution
%
%
\begin{wrapfigure}{r}{8cm}
\epsfig{figure=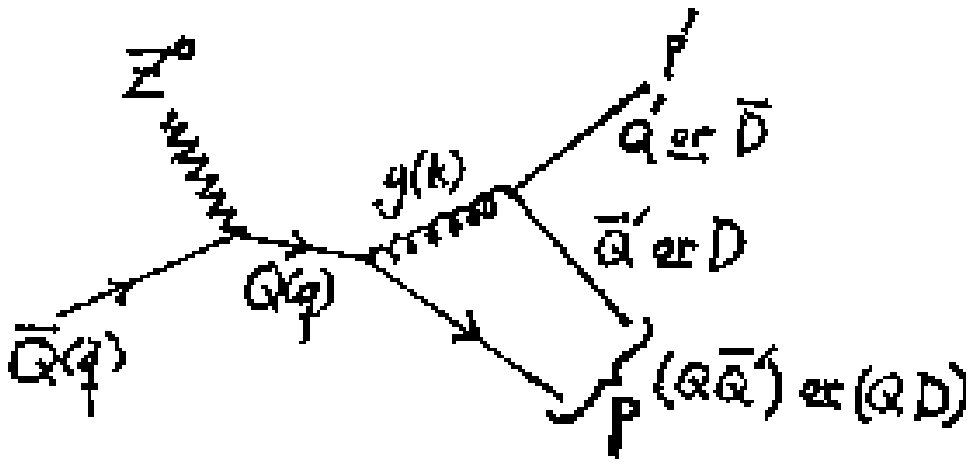,width=8cm}
{\small Figure 1: The amplitude for $Z^0 \rightarrow Meson(Q\bar{Q}')+X$
or Baryon(QD)+X.}
\end{wrapfigure}
involves the virtual b-quark
radiating a hard gluon (in axial gauge, so that there is no contribution 
from the opposite quark). The hard gluon produces a heavy flavor pair of 
c-quarks. The gluon must have energy at least twice the charm mass, so 
that the coupling $\alpha_s(Q^2)$ is small. For matching momenta 
(or relative 3-velocity zero) the $b$ and 
$\bar{c}$ form the $B_c$ meson with probability given by the square of the 
Bethe-Salpeter wave function. Since the doubly heavy mesons are weakly 
bound objects, the wave function at the origin is known from non-relativistic 
quark models for the heavy-heavy system. 

The amplitude for Fig.1, $A_1$, can be evaluated explicitly from 
perturbation theory.
The decay rate for unpolarized $Z^0 \rightarrow B_c+\bar{c}+b$ 
can be written
\begin{equation}
\Gamma_1\,=\,\frac{1}{2M_Z}\!\int\![d\bar{q}][dp][dp']\,(2\pi)^4
\delta^4(Z-\bar{q}-p-p')\frac{1}{3}\!\sum\!|A_1|^2,
\label{eq:gam1}
\end{equation}
where $\bar{q}$, $p$, and $p'$ are the 4-momenta of the $\bar{b}$, $B_c$, and
$c$, respectively.

To obtain the full inclusive width the unobserved quarks must be 
integrated over.
The phase space integration can be simplified considerably when the 
limiting case of $M_Z \rightarrow \infty$ is approached by taking leading 
order in $m_b/M_Z$ (with $m_c<m_b$). The two 
body phase space for $p$ and $p'$ can be written as an integration over z 
and s, with transverse momentum fixed for each such pair.
In the large $M_Z$ approximation the 
transverse momentum of the hadron is small and $p=zq$. Once the square of 
the amplitude $A_1$ is summed over spins and simplified by dropping 
non-leading contributions, the width for $Z^0\rightarrow \bar{c}c$ can be 
factored out of the expression Eq.~\ref{eq:general}
leaving an integral over the fragmentation function. Then
\begin{equation}
\int_0^1\!dz\,D_{c\rightarrow H}(z)=\frac{8\alpha_s^2|R(0)|^2}{27\pi m_c}
\!\int_0^{\infty}\!ds\!\int_0^1\!dz\,\Theta(s-\frac{4m_c^2}{z}-
\frac{m_b^2}{1-z})F(s,z),
\label{eq:frag}
\end{equation}
where R(0) is the Bethe-Salpeter wavefunction at the origin, and
F is the remaining integrand, which depends on $s=q^2$, z and the 
quark masses. The upper limit on the $s$ integration appears as the 
$M_Z\rightarrow\infty$ limit. 
So the partial width for $Z^0 \rightarrow H + X$ is given by an 
integral over the virtuality of the heavy quark and the phase space of the 
unobserved degrees of freedom. 

The same procedure can be applied directly to the baryons, if the 
quark-diquark model of the baryons is used. The hard gluon in the process 
must produce a diquark--anti-diquark pair, and the diquark (color 
anti-triplet) combines with the quark to form the baryon. Note that an 
alternative scenario has the heavy quark fragment into a diquark first, 
and then the diquark dresses itself to form the baryon\cite{falk}. This 
leads to very different results, as pointed out in Ref.\cite{russians}, 
and is not justifiable herein, where the diquark is not necessarily heavy 
flavor. These latter authors~\cite{russians} have performed a calculation
that is similar in spirit to part of the spin independent procedure we
follow below.

We now proceed with the calculation of
fragmentation functions for (singly) heavy flavor baryons.
The basic covariant coupling of diquarks to gluons was written long 
ago\cite{gold1}. There is one coupling constant for the scalar diquark 
color octet vector current coupling to the gluon field---a color charge 
strength, along with a possible form factor $F_s$. The momentum space color 
octet current (which couples to the gluon field vector) is
\begin{equation}
J_{\mu}^{A(S)} = g_s F_s(k^2) (p+p')_{\mu}S^{\alpha 
\dagger}\lambda_{\alpha \beta}^{A}S^{\beta},
\label{eq:scalar}
\end{equation}
where $p$ and $p'$ are the scalar diquark 4-momenta and $k=p'-p$.
For the vector diquark there 
are three constants - color charge, anomalous chromomagnetic dipole 
moment $\kappa$,
and chromoelectric quadrupole moment $\lambda$, along with the 
corresponding form factors, $F_E,\,F_M,\,{\rm and}\,F_Q$. 
\begin{equation}
\begin{array}{rcl}
J_{\mu}^{A(V)} & = & g_s(\lambda^A)_{\beta 
\alpha}\left\{F_E(k^2)[\epsilon^{\alpha}(p)
\cdot\epsilon^{\beta \dagger}(p')](p+p')_{\mu}\right.\\
   & & +(1+\kappa)F_M(k^2)[\epsilon_{\mu}^{\alpha}(p)p
\cdot\epsilon^{\beta \dagger}(p')+\epsilon_{\mu}^{\beta
\dagger}(p')p'\cdot
\epsilon^{\alpha}(p)]\\
   & &
 +\frac{\lambda}{m_D^2}F_Q(k^2)[\epsilon_{\rho}^{\alpha}(p)
\epsilon_{\nu}^{\beta \dagger}(p')+\frac{1}{2}g_{\rho \nu}
\epsilon^{\alpha}(p)\cdot\epsilon^{\beta 
\dagger}(p')]k^{\rho}k^{\nu}(p+p')_{\mu}\left.\right\},
\end{array}
\label{eq:vector}
\end{equation}
where $A$ is the color octet index, $\alpha,\,\beta$, ..., are color 
anti-triplet indices, the $\epsilon$'s are polarization 4-vectors for the
diquarks.

In the perturbative diagrams 
involved here, the virtual heavy quark emits a time-like off-shell gluon, 
that produces a diquark-antidiquark pair, while attaining nearly on-shell 
4-momentum. The diquark combines with the heavy quark to form a heavy 
flavor baryon, whose amplitude for formation is related to the 
Bethe-Salpeter wavefunction for the diquark-quark system. As in the meson 
production calculations, it 
is assumed that the constituents are heavy enough so that the binding is 
relatively weak, i.e. the quark and diquark are both on-shell and the 
binding energy is negligibly small. This is expected to be true for 
constituents with masses well above $\Lambda_{QCD}$, and even the 
light flavor diquarks almost satisfy this constraint. The basic 
perturbative amplitude is shown in Fig.1 with the Q$'$-quark line 
replaced by an (anti-)diquark D line. 

It should be realized that the integration (over s, the square of the 
virtual heavy quark mass) involved in the calculation 
would diverge for point-like vector diquarks, since the gluon coupling to 
a pair, Eq.~\ref{eq:vector} carries momentum factors. 
The virtual mass in the integration, $\sqrt{s}$, is passed on 
to the gluon and, subsequently, to the gluon-diquark vertex. 
Hence it is essential to regulate 
the integrand by some means. This is best accomplished via the 
chromoelectromagnetic form factors 
for the gluon coupling to the diquark. The 
form factor approach makes physical sense - it is a result of the 
compositeness of 
the diquarks. And for consistency, once the vector has form factors, the 
scalar diquark must have one also. 

There is no direct information about the chromoelectromagnetic form factors. 
We may expect that the ordinary electromagnetic form 
factors will have the same functional form as their QCD counterparts---the 
source of both sets of form factors is the matrix element of a 
conserved vector current 
operator. In the relevent case here, though, the vector operator is the 
gluon field --- a color octet. Also, what is of concern here is the 
time-like region of the form 
factor. For diquarks, of course, there is not any direct empirical 
evidence about their electromagnetic form factors, but diquark-quark models 
of the nucleon 
have constrained the parameterization of the form factors. Dimensional 
counting rules require that asymptotically, $F_S\sim 1/|q|^2$, $F_{E\,{\rm 
or}\,M}\sim 1/|q|^4$, and $F_Q\sim 1/|q|^6$. 
Jacob, {\it et 
al.}\cite{kroll}, have obtained electromagnetic form factors for the 
diquarks (as has a recent study of higher twist contributions to the 
nucleon structure functions\cite{anselmino}). The diquark form factors
are assumed to have simple pole or
dipole forms, and the resulting pole positions appear near 1 GeV. 
If we make the assumption that the color 
form factors have the same functional form as these empirical 
electromagnetic form factors, we can proceed. 
In the integration that 
will be done here, the time-like $q^2$ region begins at $4m_{Diquark}^2$ 
(below the $N\bar{N}$ threshold) for the value $z=1/(1+m_D/m_B)$, and at 
higher values for other choices of 
z. This implies that the integration region either overlaps or comes near 
to overlapping the pole positions. 
We treat the poles as real resonance 
positions, by including a sizeable imaginary part (of 1 GeV).
This is sensible physically, since the color octet form factor would be 
dominated by color octet vector mesons, and the latter are not expected to 
be strongly bound or narrow. Hence we have the Breit--Wigner forms and their 
squares, with pole positions as given by Jacob, {\it et al.}.
This choice hides our ignorance and provides an 
interpolation between the 
space-like and time-like asymptotic regions. 

The amplitudes for the baryon production can now be calculated. The spin 
1/2 ground state baryons are composed of a scalar diquark and a heavy 
quark in an s-state. There is only one coupling, and it involves the 
$F_S$. The amplitude is
\begin{equation}
A_{S\,1/2}=-\frac{\psi(0)}{\sqrt{2m_d}}F_S(k^2)\bar{U}_Bg_s
[k_{\lambda}-
2m_d v_{\lambda}]P^{\lambda},
\label{eq:As}
\end{equation}
where 
\begin{equation}
P^{\lambda}=
\bigtriangleup^{\lambda}_{\nu}g_s\gamma^{\nu} 
\frac{m_Q(1+\mathbf{v})+\mathbf{k}}{(s-m^2_Q)}\Gamma.
\label{eq:Plambda}
\end{equation}

For the vector diquark baryons, there are two form factors (we 
take the quadrupole to be zero -- it falls as $1/|q|^6$ asymptotically). 
The chromomagnetic coupling involves a parameter $\kappa$, the ``anomalous 
chromomagnetic moment''. This is taken as 1.39.
The s-state baryons are spin 3/2 and 1/2, which we will refer to as 1/2$'$. 
The 1/2$'$ lies between the 3/2 and the ground state 1/2 baryon. The 
amplitude for vector diquarks to be produced, along with the heavy quark, 
contributes to both 3/2 and 1/2$'$ states. The amplitude is conveniently 
divided into a chromoelectric and chromomagnetic part, involving the two 
distinct form factors. The chromoelectric part contributing to the spin 
1/2$'$ baryon is
\begin{equation}
A_{E\,1/2}=-\frac{\psi(0)}{\sqrt{3m_d}}F_E(k^2)\bar{U}_B\gamma_5\gamma^{\mu} 
\frac{1+\mathbf{v}}{2}g_s\bar{\epsilon}^{*}_{\mu}[k_{\lambda}-
2m_d v_{\lambda}]P^{\lambda}.
\label{eq:E1/2}
\end{equation}
The chromomagnetic contribution to the spin 1/2$'$ baryon is
\begin{equation}
A_{M\,1/2}=\frac{\psi(0)}{\sqrt{3m_d}}F_M(k^2)(1+\kappa) 
\bar{U}_B\gamma_5\gamma^{\mu} 
\frac{1+\mathbf{v}}{2}g_s[g_{\mu\lambda}(\bar{\epsilon}^{*}v)m_d-
\bar{\epsilon}^{*}_{\lambda}k_{\mu}]P^{\lambda}.
\label{eq:M1/2}
\end{equation}
For the spin 3/2 baryon the corresponding amplitudes are
\begin{equation}
A_{E\,3/2}=-\frac{\psi(0)}{\sqrt{2m_d}}F_E(k^2)\bar{\Psi}^{\mu}_B
g_s\bar{\epsilon}^{*}_{\mu}[k_{\lambda}-
2m_d v_{\lambda}]P^{\lambda},
\label{eq:E3/2}
\end{equation}
and
\begin{equation}
A_{M\,3/2}=\frac{\psi(0)}{\sqrt{2m_d}}F_M(k^2)(1+\kappa)\bar{\Psi}^{\mu}_B
g_s[g_{\mu\lambda}(\bar{\epsilon}^{*}v)m_d-
\bar{\epsilon}^{*}_{\lambda}k_{\mu}]P^{\lambda}.
\label{eq:M3/2}
\end{equation}
In these amplitudes, $\psi(0)$ is the Bethe-Salpeter 
wavefunction at the 
origin (for the s-state Q-diquark system), $m_d$ is the appropriate 
diquark mass, $U_B$ is a spin 1/2 Dirac spinor for the baryon, $\Psi^{\mu}_B$ 
is the Rarita-Schwinger spinor for the spin 3/2 baryon, $v=p/M$ for the 
heavy baryon of mass M, $k$ is the 4-momentum of the gluon and 
$\bigtriangleup^{\lambda}_{\nu}$ is the corresponding propagator in axial 
gauge, $\Gamma$ is the production vertex for 
the heavy quark--antiquark pair. Each
%
%
\begin{wrapfigure}{r}{8cm}
\epsfig{figure=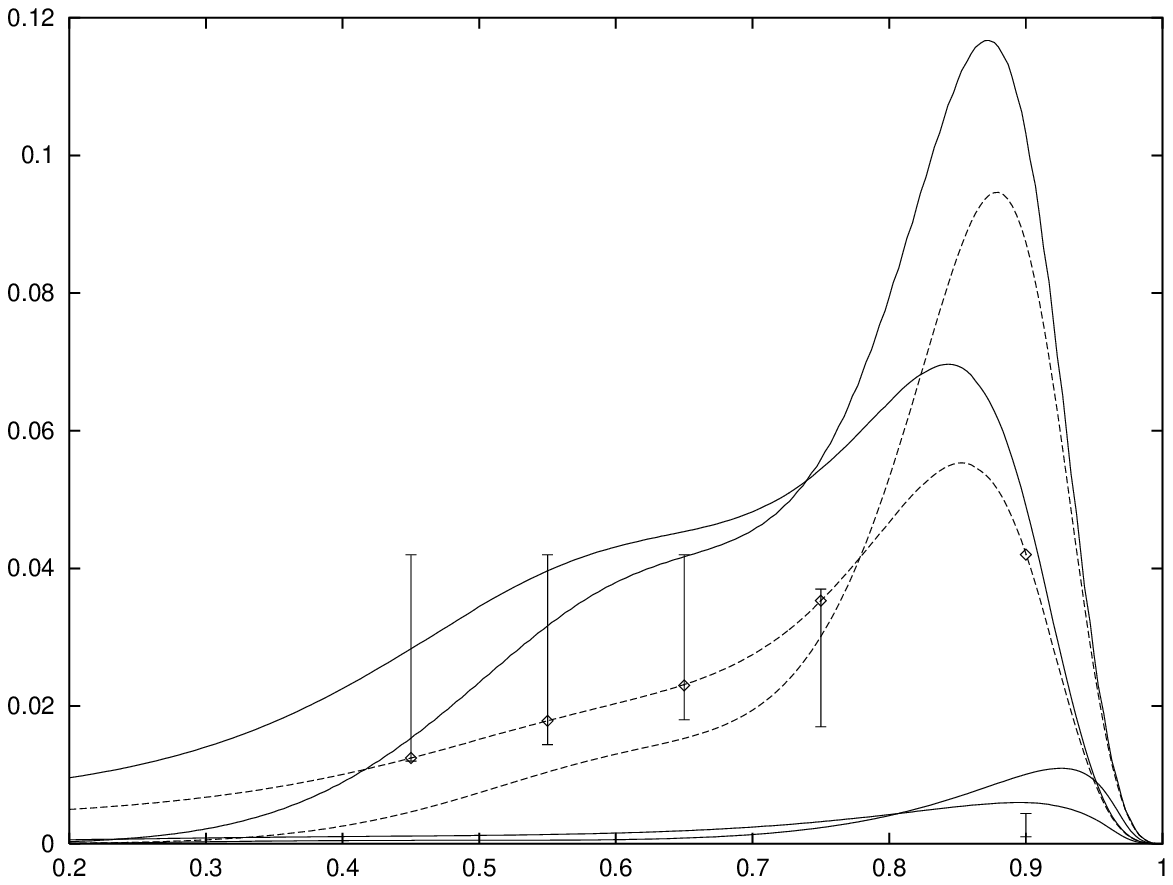,width=8cm}
{\small Figure 2:$\hat{f}_1(z,Q^2)$ for 1/2$'$ (upper pair), 3/2 (middle),
1/2 baryons, each at $Q=\mu$ and 5.5 GeV.}
\end{wrapfigure}
amplitude should be multiplied by the color factor $4/3\sqrt{3}$.

Considerable simplification of these amplitudes follows. There are 3 cases 
to consider---3 final state baryons. For the two states resulting from the
vector diquark, the electric and magnetic amplitudes must be added
together. Then for each baryon, the 
amplitude is squared and a trace is taken to sum over spins (including 
spin projection operators for the spin dependent cases). The analog of 
Eq.~\ref{eq:gam1} is obtained for each baryon. By carefully organizing the 
terms in the integrand, the width for the inclusive production of the 
virtual heavy quark can be divided out to yield the analog of 
Eq.~\ref{eq:frag} for each baryon. Finally the integration over $s=q^2$ 
can be performed numerically---the form factors make it difficult to 
write an analytic expression for each case. The resulting $z$ dependent 
fragmentation functions are the ``boundary'' functions, obtained at a
scale $\mu^2$ at the threshold $4m_D^2$. To consider higher momentum
scales, 
the Altarelli-Parisi evolution equations are used. 

We have taken some particular cases to illustrate the results. For the 
c(su) or c(sd) baryons, the $\Xi_c$ states, the diquark is given a mass of
0.95 GeV/c$^2$ and the ratio of diquark to hadron mass is $r=0.33$. The 
resulting function, $\hat{f}_1(z,Q^2)$ is shown in Fig.2 for the boundary 
value at $\mu=2m_d$ and
for $Q=5.5$ GeV. The three s-states lead to 
different behavior and overall probability. It is particularly noteworthy 
that the 1/2 ground state is produced far less frequently than the 3/2 
state or the 1/2$'$ state from the vector diquark. As we will see, the 
observed 1/2 ground states are primarily from the decays of these vector 
diquark states. 

%
%
\begin{wrapfigure}{r}{8cm}
\epsfig{figure=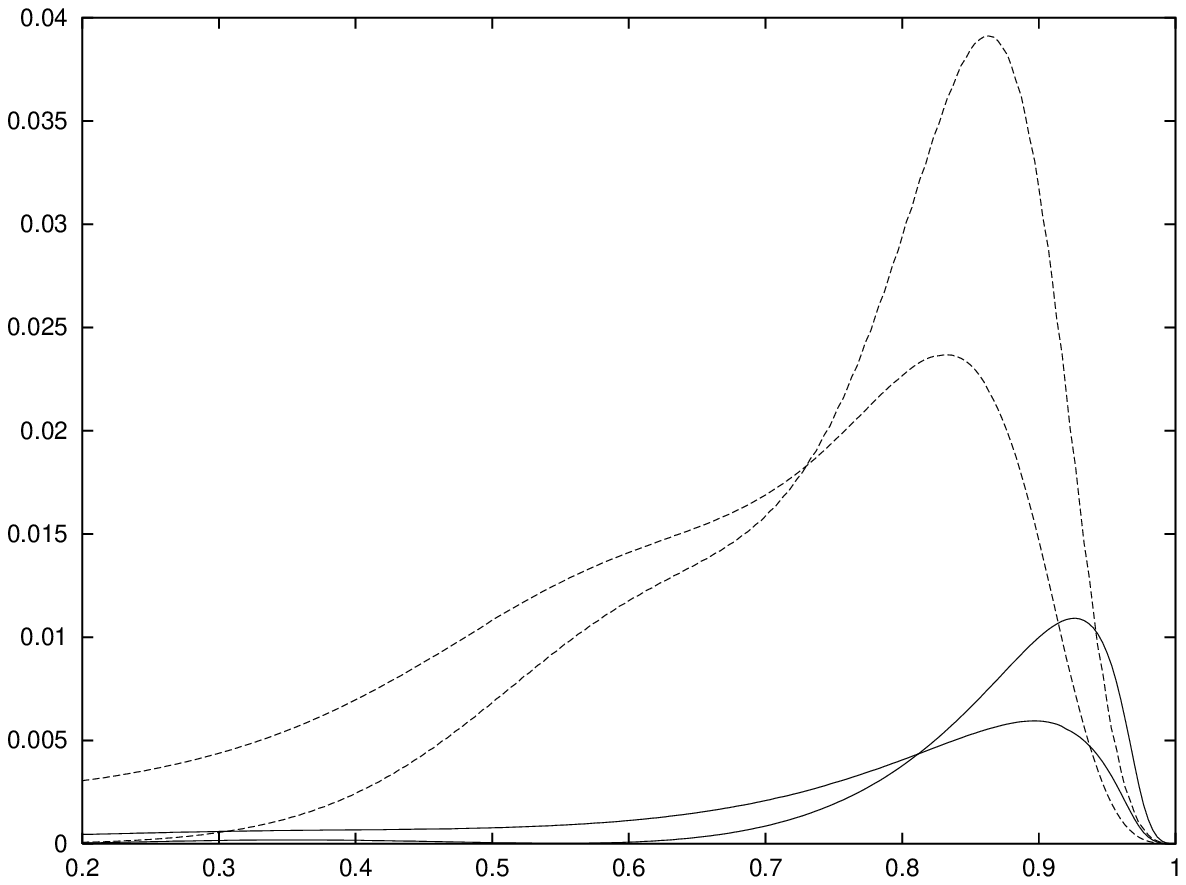,width=8cm}
{\small Figure 3:$\hat{g}_1(z,Q^2)$ for 1/2$'$ (upper pair) and 1/2
baryons, each at
$Q=\mu,\,5.5$ GeV.}
\end{wrapfigure}

In Fig.3 the corresponding longitudinal fragmentation 
function $\hat{g}_1(z,Q^2)$ is plotted for the two spin 1/2 states. Note 
that it is very similar in shape to the spin averaged case. 
Lastly, the behavior of the transversity function    
$\hat{h}_1(z,Q^2)$ remains to be determined.

The spin dependent fragmentation functions for the spin 3/2 baryons are 
even richer in complexity. There are seven such functions at leading 
twist, many of which will be accessible from the decay distributions of 
these states into the 1/2 state plus a pion. 

\section{Comparison with some data and summary}

The production of 
charmed baryons is becoming sizeable at CESR and data 
now exist from the CLEO collaboration\cite{CLEO} for some of the 
$\Xi_c$ states. In particular, spin 
independent fragmentation functions have been determined for the lowest 
mass spin 1/2 and 3/2 states. The lowest $1/2^+$ states are the $c+[u,s]$ 
and $c+[d,s]$ states, $\Xi_c^{+}$ and $\Xi_c^{0}$, 
involving the antisymmetric, spin 0 diquarks. The 
spin $3/2^+$ states are $c+\{u,s\}$ and $c+\{d,s\}$ baryons, 
$\Xi_c^{*+}$ and $\Xi_c^{*0}$, involving the 
symmetric, spin 1 diquarks. The spin $1/2^+$ partners, $\Xi_c^{'}$, of 
$3/2^+$ states 
have not been seen yet. They are presumed to have a mass below the 
$\Xi_c+\pi$ threshold, so must be seen in radiative decay channels. Note 
that these latter $\Xi_c^{'}\, 1/2^+$ states have the same isospin 
as the lower lying 
ground states $1/2^+$ $\Xi_c$ and could mix with them, in principle. In 
any case, the measured fragmentation functions provide a crude test of the 
model. The data are fit by the experimenters with a common
parameterization of the Peterson 
function\cite{peterson}. It is easy to see in Fig.2 that the data fall 
nicely on $\hat{f}_1$ of our model(with arbitrary normalization), evolved
to $Q=5.5$ GeV, but are not 
sufficiently accurate to be a crucial test of the model. Note that the
experimental variable $x_p$~\cite{CLEO} does not correspond exactly to our
$z$, the light cone variable.

The ratio of the $3/2$ to $1/2$ production can 
be extracted from 
the data with some uncertainty\cite{yelton}. The percentage of all 
$\Xi_c^{+}$ states that arose from decays $\Xi^{*0}_c \rightarrow 
\Xi_c^{+} + \pi^{-}$ is given as $(27 \pm 8)\%$ and the 
percentage of all 
$\Xi_c^{0}$ states that arose from decays $\Xi^{*+}_c \rightarrow 
\Xi_c^{0} + \pi^{+}$ is given as $(17 \pm 6)\%$. (Note that we have combined 
the statisitical and systematic errors here.) 

The experimenters do not see the $\pi^{0}$ channels, $\Xi^{*+}_c 
\rightarrow \Xi_c^{+} + \pi^{0}$ and $\Xi^{*0}_c 
\rightarrow \Xi_c^{0} + \pi^{0}$. From isospin conservation these channels 
account for 1/3 of the decays into $\Xi_c + \pi$, while the reported 
charged $\pi$ channels constitute 2/3. Suppose $N \, \Xi^{*}_c$ states of 
both charges are produced. Then 2/3 N will be seen in the charged $\pi$ 
decay mode. The total number of $\Xi_c^{+,0}$'s seen will be $N_{+,0} = 
\frac{2}{3}\!N/(0.27,0.17)$ (supressing errors until the end). The number 
of $\Xi_c^{+,0}$'s not coming from the decays of the 3/2 states will be 
$N_{+,0} - N$. Assume that $n_{+,0}$ of the $\Xi_c^{+,0}$'s come from 
other 
fragmented states' decays. Then $N_{+,0} - N - n_{+,0}$ is the number of direct 
fragmentation products of the charmed quark. The ratio $R(+\,{\rm or}\, 
0)$ of directly 
fragmented $\Xi_c^{+,0}$ to $\Xi^{*+,0}_c$ is given thereby as $R(+) = 1.5
\pm 
0.7 - n_{+}/N: 1$ and $R(0) = 2.9 \pm 1.4 - n_{0}/N : 1$. 

The numbers $n_{+,0}$ will come from the radiative decays of the heavier 
1/2 states, as well as higher $\Xi_c$ states (radial and orbital 
excitations of the $c+(su)$ and $c+(sd)$ systems). We have calculated the 
fragmentation functions for the spin 1/2 quark--vector-diquark states 
and hence the number of $\Xi_c^{'}$ spin 1/2$'$ states vs. $\Xi^{*}_c$
spin 3/2 states. That is 1.7:1 for the parameterization used in Fig.2. 
Assuming $n_{+,0}$ is due entirely to these $1/2'$ states decaying 100\% 
into the ground state $\Xi_c^{+,0}$, we have for the 
different charge states $R(+) = -0.2 \pm 0.7$ and $R(0) = 1.2 \pm 1.4$, 
both of which are consistent with the small ratio of 1/9 predicted by the 
same model calculation. 

Hence an interesting feature of our model is that the directly fragmented 
gound state baryon is very rare compared to the vector diquark states.
In the calculations for heavy-heavy baryons by Marteynenko and
Saleev~\cite{russians}, a  ratio closer to unity was obtained. It 
will be interesting to see whether the small ratio, as suggested by our
model, persists as more data are obtained.

\section*{Acknowledgments}
This work was supported, in part by a grant from the US Department of 
Energy. G.R.G. appreciates the hospitality of the organizers of 
Diquark III. We appreciate helpful communications with J. Yelton regarding 
CLEO results.

\end{document}